\documentclass{aa}  

\usepackage{siunitx}
\usepackage{graphicx}
\usepackage{txfonts}
\usepackage{algorithm}

\usepackage{natbib}
\usepackage[tight]{units}
\usepackage{nicefrac}

\bibpunct{(}{)}{;}{a}{}{,} 

\usepackage{xcolor} 
\usepackage[colorlinks=true,citecolor=blue]{hyperref}

\begin{document}

\title{3-beam self-calibrated Kernel nulling photonic interferometer}

\author{Nick Cvetojevic \inst{1} \and Frantz Martinache \inst{1} \and Peter Chingaipe \inst{1} \and Romain Laugier \inst{2} \and Katarzyna Ławniczuk \inst{3} \and \\Ronald G. Broeke \inst{3} \and Roxanne Ligi \inst{1} \and Mamadou N'Diaye \inst{1} \and David Mary \inst{1}
}

\institute{ Université Côte d'Azur, Observatoire de la Côte d'Azur, CNRS, Laboratoire Lagrange, France.
\and
Institute of Astronomy, KU Leuven, Celestijnenlaan 200D, 3001, Leuven, Belgium. 
\and
Bright Photonics B.V., Horsten 1, 5612 AX Eindhoven, the Netherlands.} 


\date{Received March 03, 2022; accepted June 00, 2022}

 

\abstract{The use of interferometric nulling for the direct characterization of extrasolar planets is an exciting prospect, but one that faces many practical challenges when deployed on telescopes. The largest limitation is the extreme sensitivity of nullers to any residual optical path differences between the incoming telescope beams even after adaptive optics or fringe-tracker correction. The recently proposed kernel-nulling architecture attempts to alleviate this by producing the destructive interference required for nulling, in a scheme whereby self-calibrated observables can be created efficiently, in effect canceling out residual atmospheric piston terms. Here we experimentally demonstrate for the first time a successful creation of self-calibrated kernel-null observables for nulling interferometry in the laboratory. We achieved this through the use of a purpose-built photonic integrated device, containing a multimode interference coupler that creates one bright, and two nulled outputs when injected with three co-phased telescope beams. The device produces the nulled outputs in a way that, by the subtraction of the measured output flux, create a single self-calibrated kernel-null. We experimentally demonstrate the extraction of kernel-nulls for up to 200~nm induced piston error using a laboratory test-bench at a wavelength of 1.55~$\mu$m. Further, we empirically demonstrate the kernel-null behaviour when injected with a binary companion analogue equivalent to a 2.32~mas separation at a contrast of $10^{-2}$, under $100$~nm RMS upstream piston residuals. }

\titlerunning{3-beam Kernel nuller}
\authorrunning{Cvetojevic et. al. }
\maketitle
%


\section{Introduction}

Most of modern optical interferometers used for astronomy utilize beam combination designed to recover the fringe visibility and phase of baselines (and their derivative observable closure-phase) made by a single or multiple telescopes.  However, another observable can be created interferometrically that is of particular interest for high-contrast astronomy, particularly exoplanet science, called nulling \citep{6_bracewell}. When conducting standard interferometry, the signal tends to be dominated by the on-axis target (typically an unresolved star), whose photon-noise can swamp the signal of any nearby faint companion. Interferometric nulling is a solution that suppresses on-axis starlight to remove the photon noise while leaving the light from any faint structure in the surrounding halo mostly unaffected, which can then be analysed separately (by sending to a spectrograph, for example). 

While astrophysical nulling interferometry using bulk-optic has a long history, including the Bracewell Infrared Nulling Cryostat (BLINC) \citep{Hinz1998}, the Keck Interferometric Nuller (KIN) \citep{KIN2}, and the Large Binocular Telescope Interferometer Nuller \citep{defrere2016}, one of the early pioneers of using photonic technologies in this domain is the Palomar Fiber Nuller (PFN) \citep{6_PFN}. The PFN instrument used the spatial filtering properties of optical fibers to improve the instrumental null, combining light from two sub-apertures of the 5.1~m Palomar Telescope at K-band. Importantly, it also introduced a novel data analysis technique of the statistical fluctuations of the null depth measurements, in essence creating a type of statistical null self-calibration (NSC) \citep{hanot2011,serabyn2019}, allowing for detection and precise measurement of faint structures.

The next major evolution was the use of photonic circuitry to create a series of pairwise Bracewell nulls inside a single integrated chip, where the dark and bright channels can be routed to a spectrograph and dispersed. This technique has most recently and comprehensively been advanced by the Guided-Light Interferometric Nulling Technology (GLINT) module \citep{norris2020,GLINT2021} on the SCExAO extreme adaptive optics instrument at the 8-meter Subaru Telescope. GLINT uses Ultrafast Laser Inscribed (ULI) 3D waveguides to sample the telescope pupil at multiple points and remap them while maintaining pathlength matching to an integrated beam-combination stage. While the first prototype was a simple 2-input combiner \citep{norris2020}, GLINT currently combines up to 6-inputs, producing 18 outputs \citep[$6\times$ pairwise Null channels, $6\times$ Bright channels, and $6\times$ photometric channels,][]{GLINT2021}. At time of writing, this is the state-of-the art for on-sky photonic nulling interferometry. 

Despite the success of nulling interferometry \citep{NullingReview2022}, most implementations thus far (both photonic and bulk-optic) have been of the pairwise Bracewell-type, and face considerable challenges when implemented on-sky. With the key requirement of nulling being the stable destructive interference of two incoming interferometer arms, a very high level of upstream wavefront control is paramount, particularly of any differential piston between the arms. While modern AO systems can correct much of this, any residual (either instrumental or atmospheric) directly translates into a photometric leakage term in the null channel, and is indistinguishable from the leakage caused by an astrophysical source. With the key scientific driver of this kind of interferometry being high-contrast detections at high angular resolution, this is highly undesired. The NSC method can be used to untangle the instrumental, atmospheric and astrophysical causes of this leakage (and retrieve only the astrophysical null) by exploiting the statistical distribution properties of the signal and various noise sources, but itself has some limitations. Firstly, the model-fitting required is fairly complex, and scales dramatically with higher number of inputs and combinations. This can potentially make it difficult to constrain binary parameters such as contrast and separation. Secondly, the need to obtain enough measurements to build accurate statistical models typically results in short integration times and large data sets, limiting observational efficiency and signal-to-noise. 

Non-nulling long-baseline and Fizeau interferometry make extensive use of the production of self-calibrated observables, like closure phases \citep{Jennison1958,BaldwinCP}, and their generalized form, kernel phases \citep{Martinache2010}, to sidestep the limitations brought by wavefront residuals. This approach has provided reliable performance at very small separations (below $\lambda / D$). Bringing together the benefits of self-calibration of interferometric observables and the photon-noise suppression of nulling is powerful and exciting prospect as it opens a hereto unexploited parameter space. The concept of creating a directly self-calibrated nulling observable has been proposed using various methods in the past, including the double-Bracewell architecture \citep{Angel1997} which can offer a form of self-calibration when implemented in a cascaded nuller \citep{Velusamy2003,KIN2}. In fact, this type of architecture has been the prime choice for space-based infrared nulling interferometer concepts (such as the Terrestrial Planet
Finder Interferometer \citep{TPFIspie2006}, DARWIN \citep{darwin}, and the more recent LIFE mission \citep{LIFE2021}), and has been demonstrated in the laboratory using bulk-optic setups in the mid-IR such as the Planet Detection Testbed \citep{Martin2012}. A different approach was later proposed by \cite{Lacour2014}, which exploits the measurement of fringes in the leakage light of the nulled channel.

Recently, \cite{Martinache2018} introduced an alternative solution, where specially designed beam combinations can produce nulled outputs whose linear combinations produce self-calibrated observables. These observables, called kernel-nulls, are robust to differential pistons (instrumentally or atmospherically induced phase error) to second order. The mathematical framework of kernel nulling has been further expanded by \cite{romain2020} and, while the work focused on a 4-input architecture for the VLTI, identified a 3-input architecture which is the simplest solution to create a valid kernel-null observable. 

In this body of work, we demonstrate the first experimental realization of one such kernel-nulling interferometer, a 3-input kernel-nuller fabricated on a photonic platform using UV photo-lithography, utilizing a Multimode Interference Coupler (MMI) component. This photonic device is used to experimentally validate the self-calibrating nature of the created nulls. Section \ref{sec:concept} outlines the fundamentals of the photonic beam combination, the MMI used, and the photonic chip. Section~\ref{sec:setup} describes the laboratory characterization setup, which allowed us to test and validate the creation of the kernel-null at multiple wavelengths across the astronomical H-band and to simulate various atmospheric residuals to demonstrate the self-calibration. In Section~\ref{sec:results} we present the  performance of the kernel-null self-calibration for different levels of input wavefront error, and demonstrate the evolution of the kernel-null when a synthetic binary signal is introduced.


\section{Photonic 3-beam Kernel nuller}
\label{sec:concept}

In this section we will cover the fundamentals of how kernel-null observables are created, and more specifically how it is achieved in the context of integrated photonic circuits. While by no means exhaustive, it provides the theoretical and practical framework used in the design and fabrication stages.


\subsection{Creating self-calibrated Kernel-null observables}
\label{sec:Kernel}

The formalism and detailed mathematical treatment on how to formulate kernel nulls has been described by \cite{Martinache2018}, and \cite{romain2020}. We will provide a quick overview limited to a 3-input interferometer. Any interferometric beam-combiner can be represented by some combiner matrix $\mathbf{M}$, essentially a complex amplitude transfer-matrix that acts on a vector $\mathbf{z}$ of discrete input electric fields to create an output electric field vector $\mathbf{x}$:
    \begin{equation}\label{eq_xMz}
        \mathbf{x} = \mathbf{M} \cdot \mathbf{z}.
    \end{equation}
In our case, $\mathbf{z}$ is a 3 element vector describing the electric field (phase and amplitude) of the light at each input. The discrete outputs of the interferometer will be some combination of the input electric fields with some internal phase offsets, expressed in $\mathbf{x}$. The nuller records the intensity/flux of the outputs (the square norm of the output electric field), $\mathbf{I} = \|\mathbf{x}\|^2$ .

If the interferometer creates nulls in a classical pairwise fashion by inducing $\pi$ phase shifts in some of the arms, the matrix $\mathbf{M}$ is real, and does not allow the formation of kernel-nulls because the output intensities it produces are a degenerate function of the target information and input phase perturbations \citep{romain2020}. If instead we allow $\mathbf{M}$ to be complex, such that two or more rows are complex conjugates of each other (i.e $\mathbf{m}_2 = \mathbf{m}^*_1 $), then a kernel-null is formed by a simple subtraction of the intensities of the corresponding two outputs:
    \begin{equation}\label{eq_km}
        \kappa(\mathbf{z}) = 
        |\mathbf{m}_1 \mathbf{z}|^2 - 
        |\mathbf{m}_2 \mathbf{z}|^2 = |\mathbf{m}_1 \mathbf{z}|^2 - 
    |\mathbf{m}^*_1 \mathbf{z}|^2 .
    \end{equation}
Provided the piston errors are small ($\sim \pm 1$~rad), the otherwise dominant quadratic error term is cancelled \citep{Martinache2018}. This shows that the subtraction of measured intensity of complex conjugate pairs of nulled outputs always produces a kernel null that is robust to arbitrary phase excursions.

\begin{figure}[htp]
\centering
\includegraphics[width=0.42\textwidth]{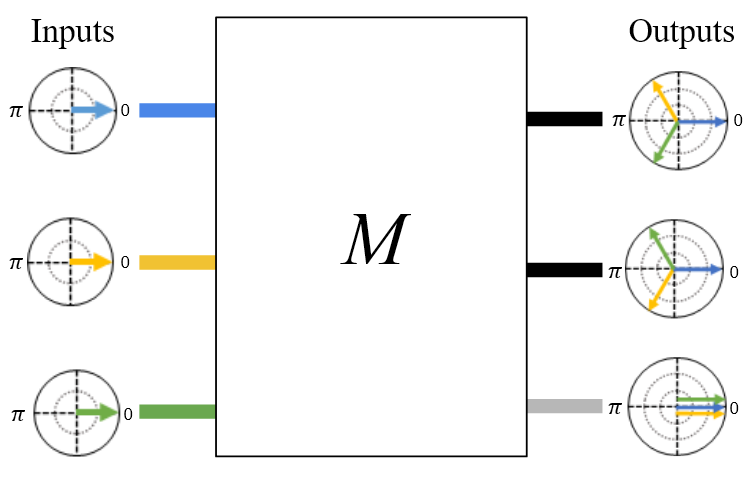}\caption{ The electric field is represented by colored phasors on a complex polar plot, with the length of the vector being the amplitude and the angle showing the phase. For three in phase inputs, the transfer matrix $\mathbf{M}$ creates three outputs. One bright channel where the input E-fields are in phase creating constructive interference, and two nulled channels where the input beams are given a $\pm 2\pi/3$ phase offset before being combined. In these channels the E-fields interfere destructively creating a null. }
\label{fig:kernel3x3}%
\end{figure}

For a 3-input interferometer the simplest $\mathbf{M}$ one can construct to form such kernel null is:
\begin{equation}\label{eq_3T_matrix}
\mathbf{M} = 
\frac{1}{\sqrt{3}}\begin{bmatrix}1 & 1 & 1 &\\
            1 &
            e^{\frac{2 j \pi}{3}} & e^{\frac{4 j \pi}{3}}\\
            1 & e^{\frac{4 j \pi}{3}} & e^{\frac{2 j \pi}{3}}\end{bmatrix}.
\end{equation}
It will result in one bright and two nulled outputs, which will (by subtraction) produce one self-calibrated kernel-null observable. In practise, this architecture is actually quite simple to create. The light from each of the 3 input beams is divided evenly and all interfered in such a way that on one output the relative phase offset is $0$ (i.e. constructive interference/ bright channel). For the other two outputs, two of the input arms are given a $\pm 2\pi/3$ phase offset. This is visualised in Figure \ref{fig:kernel3x3}.


\subsection{Multimode interference couplers}
\label{sec:MMIs}

\begin{figure}[htp!]
\centering
\includegraphics[width=0.45\textwidth]{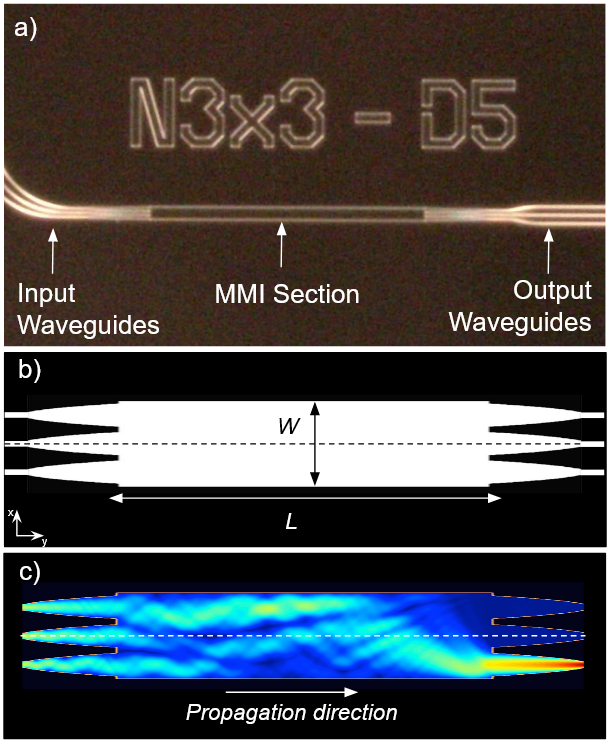}
\caption{ A microscope image (a) of the 3-input MMI component used in this work, that produces the phase-shifts and beam-combinations required by $\mathbf{M}$. To produce a desired electric field transfer from input to output waveguides, the length ($\mathbf{L}$) and width ($\mathbf{W}$) of the MMI section is carefully chosen (b). To decrease internal losses the waveguides are connected to the MMI section by a small parabolic taper region to maximize mode-matching. A beam-propagation model (c) shows how the multiple inputs interfere inside the MMI section producing a single bright output and two nulled outputs.}
\label{fig:MMI}%
\end{figure}

While it is possible to envision a bulk-optic interferometer design that creates the beam-combinations required by $\mathbf{M}$ \citep{Guyon2013}, it is far more advantageous to exploit photonic technologies to achieve the same goal. In fact, it is possible to create $\mathbf{M}$ using a single stand-alone component, in common use for a wide range of photonic designs. This component is a Multimode Interference Coupler (MMI). An MMI is a planar multimode waveguide section (often rectangular) where multiple incoming waveguides are allowed to interfere spatially to form the required intensity profile at the MMI end face (Fig. \ref{fig:MMI}a). There, output waveguides are placed that will discretely sample this intensity profile and route it to the chip output. MMIs have been a standard component in photonics for the last few decades, their behaviour, properties, and design methodology are well understood in the literature for most photonic platforms \citep{MMI_first, MMI_review}. 

For the purposes of designing a kernel nuller, the transfer matrix of the MMI itself can be tuned using a number of parameters. The most important are the length ($\mathbf{L}$) and width ($\mathbf{W}$) of the multimode section where the interference takes place, as well as the position of the input waveguides with respect to the MMI section. The last important parameter is the positions of the output waveguides which discretely sample the complex electric field. By tuning these four parameters it is possible to design an MMI that closely matches $\mathbf{M}$ (Fig. \ref{fig:MMI}b). For the MMI used in this work, the parameters were; $\mathbf{L} = 330$~$\mu$m, $\mathbf{W} = 15$~$\mu$m, and the input and output waveguide positions of $0, \pm 5~\mu$m. An additional feature typically found in MMIs are input and output adiabatic taper regions at the waveguide-MMI interface. These are employed to reduce losses due to modal miss-match between the fundamental waveguide mode and the multimode section. The tapers are essentially a broadening of the waveguide core over a short length, allowing the fundamental mode to expand. For the tapers in our MMI, they expand from the waveguide core diameter (1.1~$\mu$m) to 4~$\mu$m, over 83.3~$\mu$m propagation length. 

It should be noted that the exact same beam combination can potentially be made using a waveguide tri-coupler \citep{tricoupler_MAM}, a 3-input version of the evanescent-field combiners (3dB couplers) used in most astronomical photonic interferometers \citep{tricoupler_aus}. While in the 3-input case there is little difference between the two approaches, we found that controlling fabrication tolerance over 10's or 100's of microns was easier than the often sub-micron precision required to make traditional couplers accurate enough for kernel nulling. Furthermore, when scaling to higher number of inputs, MMIs designs are far more accommodating \citep{MMI_review}.


\subsection{The Photonic Device}
\label{sec:photonic chip}

\begin{figure}[htp!]
\centering
\includegraphics[width=0.38\textwidth]{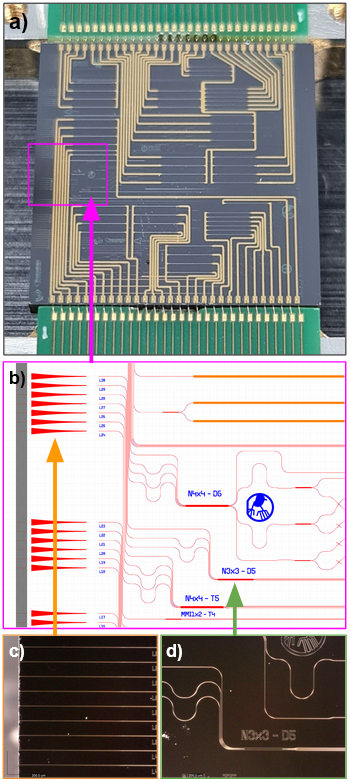}
\caption{ The photonic chip made in Silicon Nitride (a) that contains the 3-input kernel nuller measures $16 \times 16$~mm in size, and contains other more complex components not covered in this work, including electrodes that are visible for on-chip phase control. The waveguide circuitry (b) includes input and output tapers or ``mode-converters'' (c) that transform a standard $\sim 8$~$\mu$m mode at the chip input (left) into a $\sim 1$~$\mu$m fundamental mode of the waveguides. The light is then routed downwards along an ``optical bus'' to create a side-step of $\sim 1-2$~mm so any uncoupled light in the bulk doesn't re-interfere and cause cross-coupling. Prior to entering the MMI, optical pathlength-matching is performed (d) to ensure all the inputs travel the same distance to the MMI entrance. The MMI output waveguides are then routed to the chip end-face on the right hand side, and can be coupled into a fiber array (or directly imaged).      }
\label{fig:chip}%
\end{figure}

The MMI component that forms the interferometer is part of a larger photonic chip fabricated to test this novel type of interferometry. The entire $16 \times 16$~mm chip is shown in Figure \ref{fig:chip}a. At this scale the waveguides are barely visible, with the prominent features being surface electrodes for active on-chip phase control, which will not be discussed in this work. The mask layout and waveguide routing was generated using Nazca Design software\footnote{https://nazca-design.org/}. The chip was fabricated using a Silicon Nitride (SiN) platform (LioniX - TriPleX\texttrademark), which creates waveguides of $\sim 1$~$\mu$m single-mode core size for the design central wavelength of $1.55$~$\mu$m. Because of the high mode confinement of this platform, tight bends can be used with low loss allowing for more complex waveguide routing (than traditional Silica-on-Silicon platforms) and denser packing of components. The drawback is that coupling directly into a 1~$\mu$m waveguide core would typically cause high insertion loss, thus all input and output waveguides on the chip are terminated with an on-chip taper (or mode-converter) which allows the direct coupling of a standard 8~$\mu$m mode (from a commercial SMF-28e optical fiber) into the waveguide with $\sim 60\%$ efficiency (Fig \ref{fig:chip}c). This ensures the light can be injected into the chip using the same methods as all other astrophotonic interferometers operating at NIR wavelengths.  

After injection, the waveguides are routed vertically by a few millimeters to create a lateral offset from the input and `side-step' the cone of uncoupled light in the substrate. This allows us to avoid many of the negative cross-coupling effects which has been shown to degrade interferometric performance in an astronomy context \citep{dfly2021}. Just prior to entering the MMI outlined in Section~\ref{sec:MMIs}, the waveguides are given extra routing to ensure the optical pathlength experienced by all the inputs is the same, and does not need to be compensated for outside the chip. After the interference is performed inside the MMI, the output waveguides are routed to the chip end-face on the far right-hand side, where another set of waveguide tapers returns the mode back to a size that is matched to standard commercial optical fiber (SMF-28e) and can be directly collected. 

Importantly, while this work was conducted on the SiN platform (for cost and prototyping lead time considerations), all components, circuits, and features described here are possible using most photonic platforms, and therefore transposable to other bandpasses.


\section{Experimental Setup}
\label{sec:setup}

\begin{figure}[htp!]
\centering
\includegraphics[width=0.37\textwidth]{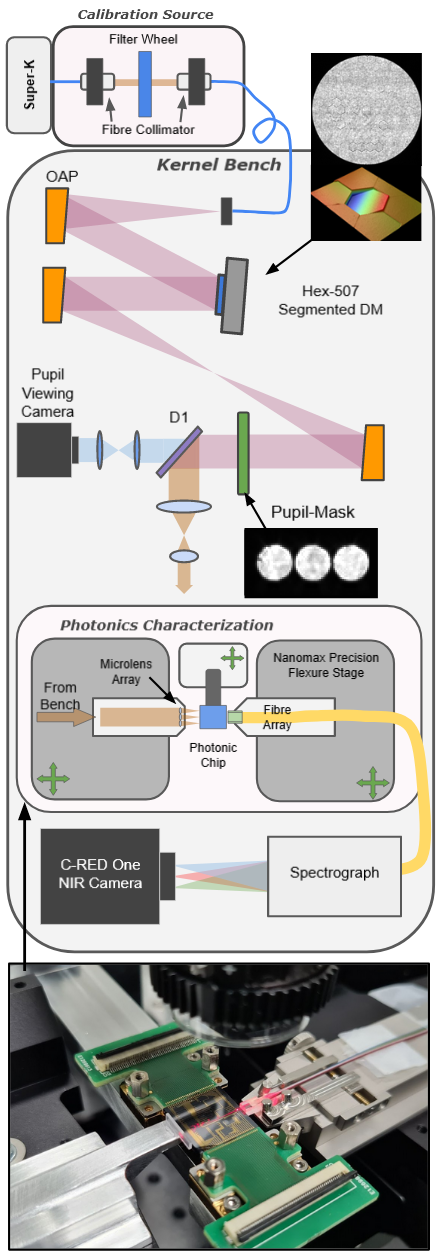}
\caption{Schematic of the experimental bench used to characterize the kernel nulling chip. A picture (bottom) shows the physical chip, the injection MLA, and output fiber array aligned using precision flexure translation stages.}
\label{fig:setup}%
\end{figure}

The 3-input kernel nuller was characterized and tested using a photonic characterization bench in an ISO-7 clean room. The bench (shown in Figure \ref{fig:setup}) consists of three major parts; accurate and fast wavefront control across a segmented pupil, precise and stable alignment of the photonic chip, and spectral dispersion onto a NIR detector. A supercontinuum source (Fianium WhiteLase micro) is used to inject a broadband light source into the bench using an endlessly single-mode Photonic Crystal Fiber (PCF) that acts as a spatially-filtered broadband on-axis point-source analogue for our experiments. Wavefront modulation, both for characterization scans and the introduction of atmospheric perturbation, relies on a hexagonal segmented (Boston Micromachines Hex-507) deformable mirror (DM), located in a plane that is conjugated with the pupil of the setup. Individual segments are controlled independently in Piston, Tip, and Tilt. With the DM in its default mid-range position, each segment had an effective maximum stroke of $\pm 2$~$\mu$m and max tip-tilt range of $\pm 3$~mrad, with a modulation speed of $> 25$~kHz. 

Only a small part of the active aperture is used in these experiments, with a brass laser-cut mask consisting of three 620~$\mu$m sized holes used to block all but three DM segments. The mask is conjugated with the pupil-plane and positioned over the segments using stepper actuators, monitored by a pupil-viewing camera at visible wavelengths. A 1~$\mu$m long-pass dichroic (D1 in Fig. \ref{fig:setup}) is used to split the NIR light and direct it to a beam-reducing telescope. Here, the beam is reduced by a factor of 5 using a pair of NIR achromatic doublet lenses, converting the mask-hole separation from 647.5~$\mu$m to 129.5~$\mu$m, close to the injection waveguide pitch of 127~$\mu$m on the chip. The telescope forms the now demagnified pupil-plane near the center of the photonic characterisation setup, where a fused silica 127~$\mu$m-pitch microlens array (MLA) was placed.

The microscopic scale of the photonics required careful, stable, and precise alignment. Thus we use a pair of high-precision Nanomax 6-axis flexure stages to align the MLA and the output fiber array to the photonic chip, which itself is on a 3-axis precision translation stage. The pupil from the bench is formed at a specific point (close to the chip's entrance face), and typically isn't moved. Instead the MLA is positioned so that each of the three (now 124~$\mu$m diameter) mask-holes overlaps a single lenslet. This is conjugated to the pupil plane by imaging the MLA front-face and pupil simultaneously using an alignment camera. Once the MLA is aligned, three focal spots are formed at the chip's input, and the chip translated so they land at the correct input waveguide positions. After the light propagates through the chip and is brought to the output face, a linear fiber array is aligned to the correct output waveguides, capturing the light into individual fibers to be routed to the spectrograph. 

The fibers coming from the photonics characterization setup are taken to a spectrograph, made from two 2-inch achromatic doublets and a F2 dispersive prism, which spectrally disperses the fiber outputs and forms an image on a NIR detector (First Light, C-RED One). The spectrograph is wavelength calibrated by referencing the various wavelength filters in a filter wheel upstream. The spectra are measured across a spectral range of $\lambda = 1.20-1.75$~$\mu$m, with a typical resolution of $ \Delta \lambda = 5.2$~nm and a nominal linear dispersion of $2.38$~nm/pix. The detector output is kept at a constant readout of 1.6~kHz, unless otherwise indicated.

\subsection{Injection Optimization}

\begin{figure}[htp!]
\centering
\includegraphics[width=0.49\textwidth]{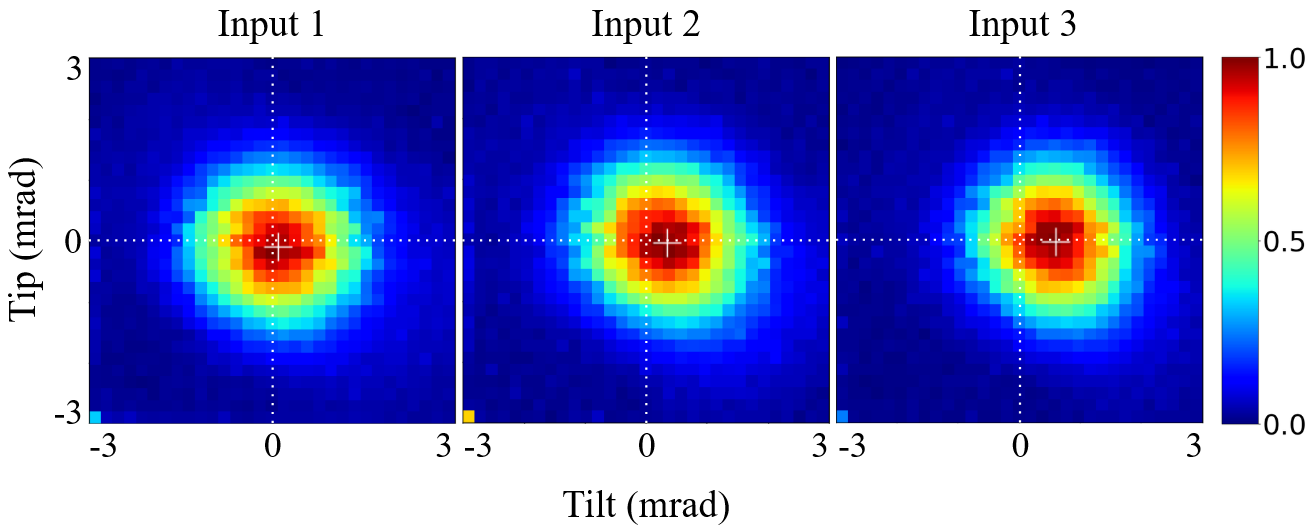}\caption{An injection optimization routine is typically done prior to acquiring nulling data to compensate for any sub-micrometer drift in the alignment. }
\label{fig:opti}%
\end{figure}

Despite the sub-micron precision of much of the alignment setup, it is near-impossible to perfectly align the chip input by hand. However, by using the tip-tilt control of each of the three active DM segments we are able to individually raster-scan the three MLA focal spots across the chip end face and find the optimal coupling location for each input to sub-micrometer precision. The typical injection optimization loop scans each input $\pm 3$~mrad with a step size of 0.2~mrad measuring the total flux of the output at every frame. After the scan is complete, a 2D Gaussian fit locates the central tip and tilt position for each input. These are then stored by the software to ensure optimal coupling for all measurements. A typical result of this kind of scan is shown in Figure \ref{fig:opti}. 

During the course of 24 hours we measure a drift of up to 1~mrad when re-optimized. This slow drift is due to the laboratory air conditioning cycle that affects the translation stage holding the chip mount and needs to be corrected before any new set of measurements. To ensure measurement consistency, the injection optimization procedure is therefore run prior to every large experimental cycle. The scan and correction procedure takes on average $\sim 4$ seconds to complete. 

\subsection{OPD zeroing}
\label{sec:opdzero}

\begin{figure*}[htp!]
\centering
\includegraphics[width=0.9\textwidth]{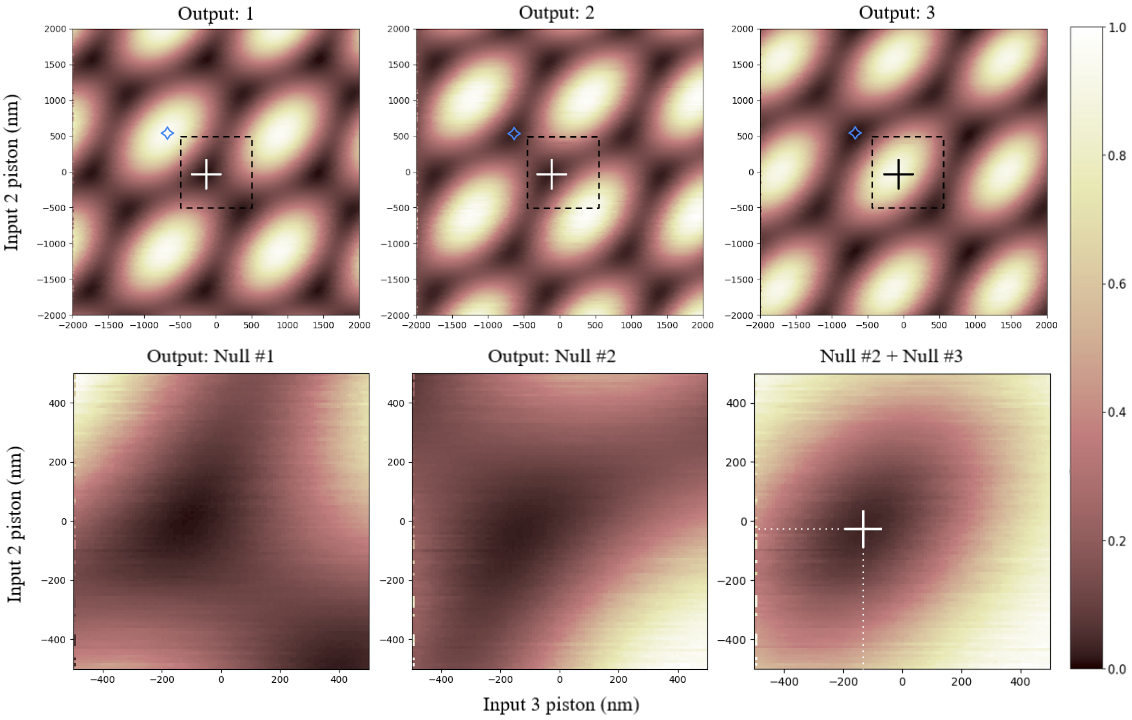}
\caption{ A scan of input pistons for input 2 (vertical) and 3 (horizontal) over at a resolution of 20~nm over a range of 4~$\mu$m (top). After a null location is chosen, a more detailed scan with a resolution of 5~mn over a 1~$\mu$m range (dashed squares) is done to find the location of the deepest raw null (bottom). The value of the two piston positions at this location ($-135$~nm~\&~$-30$~nm - white cross) is saved and used as a zero-point for all measurements. The color intensity is normalised to the maximum flux measured in the bright channel over the full scan. For reference, the blue star indicates the piston offsets used when simulating a binary companion later in Section \ref{sec:results_companion}.}
\label{fig:kscan}
\end{figure*}

The response of the nuller is a sensitive function of the OPD affecting the individual input beams. These OPDs can either come from the bulk optics prior to injecting into the device or manufacturing imperfections in the waveguides inside the device, before entering the MMI component. The experimental setup allows us to compensate for these imperfections using the upstream segmented mirror. With only three inputs, we can afford to do an exhaustive piston scan for two input beams, relative to the third, left untouched, and used as a reference.

The top panel of Fig.~\ref{fig:kscan} shows what one such scan looks like over a wide 4~$\mu$m range piston scan ($\pm$1~$\mu$m modulation of the DM segments). This first experimental result confirms that the monochromatic response is \(\lambda\)-periodic and repeats every $1550$~nm. The set-point of the nuller is, by design, where the light of the first two outputs are the darkest and the third is the brightest. Since the starting point of the DM is supposed to be flat, differential OPDs greater than \(\lambda\) can be excluded, the selected set-point is the closest to the origin of the scan. A higher resolution ($5$~nm) scan (bottom panel of Fig. \ref{fig:kscan}) is used to better constrain the ($-135$~nm, $-30$~nm) OPD offset that brings the device to its set-point.

Much like the injection optimization we don't observe a substantial drift in these values over short timescales, however over the course of a day the OPD zero can drift by 10's or 100's of nanometers, due to the room air conditioning cycle and upstream misalignment. This scan is therefore repeated before every set of measurements presented here, and takes on an additional $\sim 5$ seconds to complete.


\section{Results}
\label{sec:results}

To demonstrate experimentally how the kernel null self-calibration works, we conducted a series of different measurements using the setup outlined in Section~\ref{sec:setup}. These can be broadly arranged into three categories; locating and measuring the instrumental nulls (Sec.~\ref{sec:instrumentalnull}), the demonstration of self-calibration under different induced wavefront errors for narrow-band operation (Sec.~\ref{sec:results_kernel}), and simulating companion detections at high angular resolution for multi aperture long-baseline interferometer (Sec.~\ref{sec:results_companion}). The latter section, while not a rigorous instrumental analogue, provides a rough example of how typical self-calibrated kernel-null data looks when observing a binary source under $100$~nm piston residuals.

The primary aim of this work is to outline the self-calibration experiments, and experimentally validate the principle of kernel-nulling. While MMIs in general are a standard photonic component, the detailed characterization of the MMIs fabricated on our chip  will be presented in a separate manuscript \citep{PeteSPIE2022}. There are however some basic properties that are important for this work. The splitting ratio, the fraction of the flux that makes it from a single input to the outputs, was measured to be $33~\pm 5~\%$ for all inputs at the central wavelength of 1.55~$\mu$m. The component-level throughput was measured to be $90 \pm 10\%$. This has a considerable measurement error as it is achieved by comparing the total MMI output flux with that of a straight waveguide section located at the extreme edge of the chip, and is highly sensitive to alignment error. Similarly, due to the difficulty in quantifying injection efficiency with our setup, the global chip throughput is hard to measure with high accuracy. Based on the characterization of similar chips \citep{lionix}, we estimate the propagation loss to be $<0.1$~dB/cm, the MMI loss $<0.5$~dB, and the input and output mode converter to be $\sim 1.5$~dB per facet. Thus an estimated end-to-end throughput of $\sim 44~\%$, including coupling losses.


\subsection{Instrumental null}
\label{sec:instrumentalnull}

At its set-point, the nuller minimizes the amount of on-axis starlight that makes it down its dark outputs. The design is of course to have a perfect extinction yet manufacturing imperfections lead to some amount of light leakage, and therefore a finite extinction. In addition to imperfections of the device itself, laboratory seeing as well as readout noise are significant contributions to the amount of light measured on the dark outputs at any instant. The combined effect of these contributions leads to generally non-trivially distributed null values \citep{hanot2011}, when operating in realistic conditions. With an estimated piston RMS less than $20$~nm, the laboratory seeing is low enough for the piston error to not be the dominant source of noise, hence we are able to measure null values that are normally distributed. We will focus on a $12$~nm wavelength bandpass (corresponding to a 4-pixel bin for our spectrograph) centered on $\lambda_0 = 1.55$~$\mu$m, for which the device was specified. This was achieved by placing a narrow-band transmission wavelength filter, with a Full Width Half Maximum (FWHM) bandwidth of 12~nm, in the injection beam using the upstream filter wheel shown in Fig. \ref{fig:setup}. 

\begin{figure}[htp!]
\centering
\includegraphics[width=0.49\textwidth]{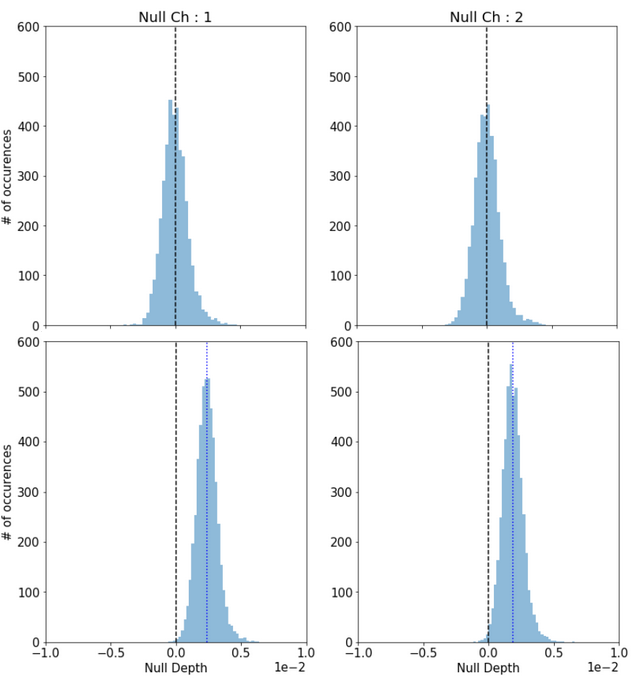}
\caption{Measured empirical distribution of the two Null channels for read-noise only (top), and the instrumental leakage (bottom) for 10,000 frames. The statistical mean of the leakage terms (blue line) is $2.4\times 10^{-3}$ and $1.8\times 10^{-3}$ for channel 1 and 2 respectively. }
\label{fig:rawnull}%
\end{figure}

At each measurement frame the photometry of all three outputs are recorded in camera counts. The photometry of the two null channels are then converted from counts into units of contrast by dividing by the total summed flux (including the bright channel). To estimate the effect of the camera readout noise (correlated double sample readout mode), the different outputs are recorded with no input light. The distribution of the readout noise (see top panel of Fig.~\ref{fig:rawnull}), built from the analysis of 10,000 individual frames, is centered on zero after dark subtraction, and is characterized by a standard deviation $\sigma_{RN} = 6.32\times10^{-4}~\pm~6.3\times10^{-6}$.

With the device brought to its set-point, light is injected, and outputs are recorded for a new series of 10,000 frames. The resulting distributions are represented in the bottom panel of Fig.~\ref{fig:rawnull}. We can observe that despite the same dark subtraction, the position of the distribution are shifted, while the standard deviation (dominated by the effect of the same camera readout noise and seeing) remains stable. We use the mean of these shifted distributions to estimate the raw instrumental null values: $n_1~=~2.397\times10^{-3}~\pm~6\times10^{-6} $ (for channel \#1) and $n_2~=~1.849\times10^{-3}~\pm~6\times10^{-6}$ (for channel \#2).

\subsection{Bench drift and measurement uncertainty}
\label{sec:drift}

While unfortunate, the $5.4\times10^{-4}$ imbalance between the two dark outputs is a feature caused by two physical parameters. First is the flux imbalance in the specific MMI used for these experiments that we have to live with, but is entirely static. The second is a differential loss term between the two channels caused by a misalignment at the chip output waveguide to fiber array interface in our setup. In any practical implementation of these devices, the fibre array would be bonded to the chip, making the coupling loss term static as well, however due to the nature of our experimental setup this is not entirely the case here. 

While the statistical uncertainty in the measurement of $n_1~\&~n_2$ are at the $10^{-5}$ level for an individual experiment, we find that between measurements it can fluctuate by up to $\sim 5\times 10^{-4}$. This is primarily due to a drift in the output fiber alignment causing a differential coupling loss change of $<1\%$ at the chip output over a 10~minute time frame. Additionally, the output flux from the super continuum source has been measured to fluctuate by $>20\%$. This typically has no effect on the null measurement, but as we are utilizing most of the dynamic range of the camera (to measure the bright and nulled channels simultaneously), a small change in total flux can cause saturation. Because the bench drifts dominate our uncertainties, all measurements presented have been repeated 10 times to derive experimental uncertainties.


\subsection{Kernel-null}
\label{sec:results_kernel}

Compared to the laboratory environment, on-sky nulling observations experience degraded seeing conditions. Post adaptive-optics and fringe tracking wavefront residuals result in additional random OPDs that pull the nuller away from the set-point and result in additional light leakage down the dark outputs. In the case of the VLTI, in the near infrared (NIR) OPD residuals can reach $\sim 100$~nm for the ATs, and $\sim 200$~nm for the UTs \citep{gravityFT}. It is this reality that motivated the kernel-nuller architecture, designed so that at any instant, simultaneously recorded values for the dark outputs can be combined to form kernel-nulls, disentangling genuine astrophysical signals from seeing induced light leakage in the different dark outputs.

\begin{figure}[htp!]
\centering
\includegraphics[width=0.49\textwidth]{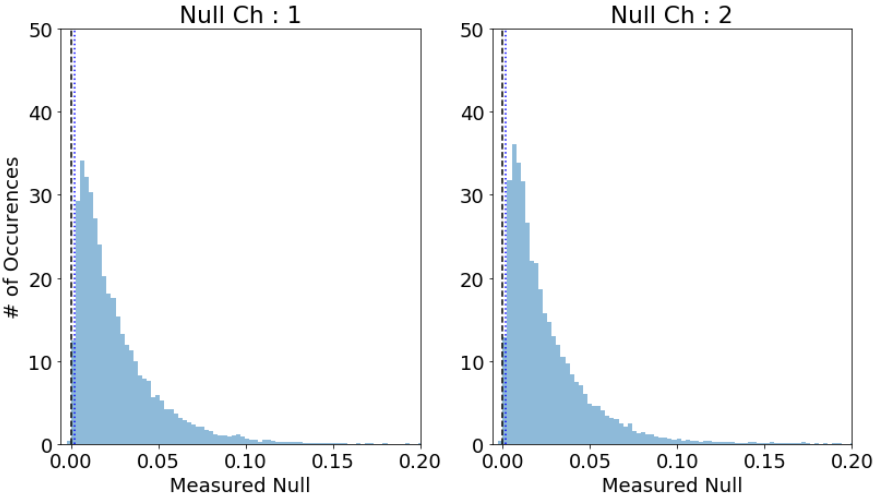}
\caption{ The statistical distribution of measured null depth in the two null channels under $100$~nm RMS wavefront error for an unresolved on-axis target. The instrumental null (blue dashed line) is visibly far away from the distribution mean, which is heavily biased by the random piston terms. Complex model-fitting analysis methods such as NSC need to be used to retrieve the true null depth.}
\label{fig:badnull}%
\end{figure}

On top of the static OPD offset applied to bring the nuller to its set-point, our setup can use the upstream DM to apply random errors, and observe the degradation of the nulled outputs as well as the benefit brought by the kernel-nuller architecture. These induced OPD fluctuations are directly applied to the input beams by the corresponding DM segments. Figure~\ref{fig:badnull}  presents the distributions of the different values recorded by the two dark outputs (20,000 frames) when the input beams experience OPD fluctuations drawn from a normal distribution with a $100$~nm standard deviation. The shape of these distributions is similar to those reported in recent on-sky uses of nulling by \citet{norris2020} and \citet{GLINT2021}.

\begin{figure}[htp!]
\centering
\includegraphics[width=0.47\textwidth]{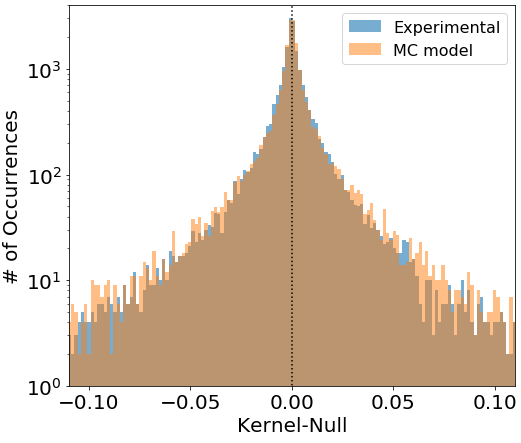}
\caption{ The statistical distribution of the measured kernel-null using the same data as in Fig.~\ref{fig:badnull}. Because the observable is self-calibrated, for an unresolved on-axis source, the statistical distribution of the observable is symmetric about the median. The median of the data set $\kappa = -6.5\times 10^{-5}$ is shown as the dotted line. The Monte Carlo simulations of an idealized 3-input kernel-nuller under 100~nm RMS residual piston show a close fit.}
\label{fig:100nmkernel}
\end{figure}

Section \ref{sec:Kernel} reminds how, with a kernel-nuller architecture, the theoretical difference between simultaneous outputs (see Eq.~\ref{eq_km}) is expected to cancel out the OPD introduced bias of the two dark outputs. At any instant $t$, for our real $3\times3$ MMI device the combination of outputs that will result in a kernel distributed as simulated by \citet{Martinache2018} is:
    \begin{equation}\label{eq_kernelmean}
        \kappa(t) = n_1(t) - n_2(t) 
    \end{equation}
Using the data from the original 20,000 frame series whose distributions are plotted in Fig.~\ref{fig:badnull}, we compute kernel-nulls using the definition of Eq.~\ref{eq_kernelmean} and plot their distribution in Fig.~\ref{fig:100nmkernel}. Whereas the interpretation of raw null distributions requires the careful modeling procedure introduced by \citet{hanot2011}, the experimental kernel-null distribution can be interpreted directly. The distribution is symmetric (see Fig.~\ref{fig:100nmkernel}) albeit non Gaussian: it features a strong narrow peak and longer tails. The characterization of the properties of this non-trivial distribution may be the object of future work, but for this work, we will rely on Monte Carlo (MC) simulations of an idealized $3\times3$ kernel-nuller experiencing phase noise that results in kernel-null distributions matching our actual experimental results (see Fig.~\ref{fig:100nmkernel}). We will use the experimental median as the estimator for the position parameter $\kappa_0$ of the underlying distribution. While the raw statistical uncertainty of a stand-alone data set is typically at a $10^{-5}$ level, it does not take into account possible drifts in the characterization setup between measurement cycles. Hence, to not overstate the precision of our bench, the associated experimental uncertainty $\sigma_\kappa$ is measured by taking the standard deviation of 10 repeat measurements with identical settings a few minutes apart. With a 20,000 frame data-set, in the presence of 100~nm RMS residual piston, the experimental kernel-null is $\kappa = -6.5\times10^{-5}~\pm~1.2\times10^{-4}$. As this data was taken separately to the raw-null measurements presented in Section \ref{sec:instrumentalnull}, the imbalance in the two null channels does not exactly equal the measured $\kappa$, due to the bench drift outlined in Section \ref{sec:drift}.

\begin{figure}[htp!]
\centering
\includegraphics[width=0.45\textwidth]{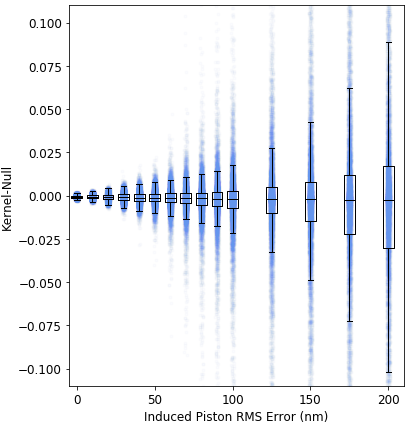}
\caption{ The measured kernel-null distributions as a function of induced piston RMS error. The box heights are defined by the first and third quartiles of each data set (encompassing 50\% of the samples), with the line at the measured median ($\kappa$). The whiskers extend from the box by one and a half times the inter-quartile range ($\sim \pm2.7 \sigma$). The raw instantaneous measurements for each frame are shown in blue. Larger piston errors broaden the distribution while remaining symmetrical.}
\label{fig:KNvsRMS}%
\end{figure}

\begin{figure}[htp!]
\centering
\includegraphics[width=0.42\textwidth]{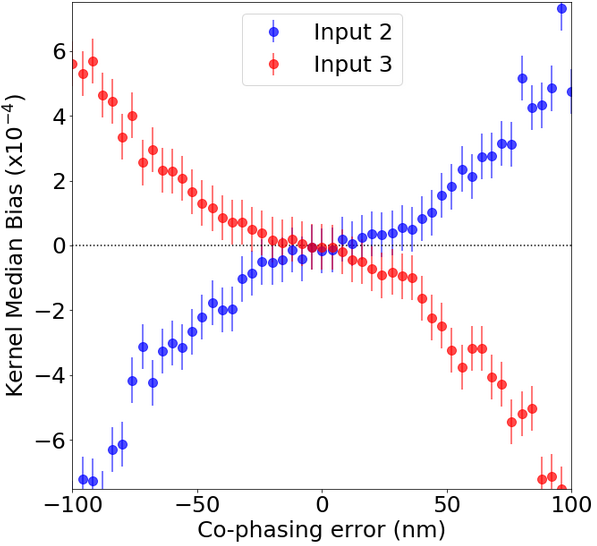}
\caption{Monte Carlo simulation showing the effect of incorrect co-phasing of input beams on the measured kernel-null median ($\kappa$) under 100~nm RMS induced piston error. The error bars represent the statistical uncertainty taken from $1000$ simulation iterations per data point.} 
\label{fig:kappabias}%
\end{figure}

\begin{figure*}[htp!]
\centering
\includegraphics[width=0.9\textwidth]{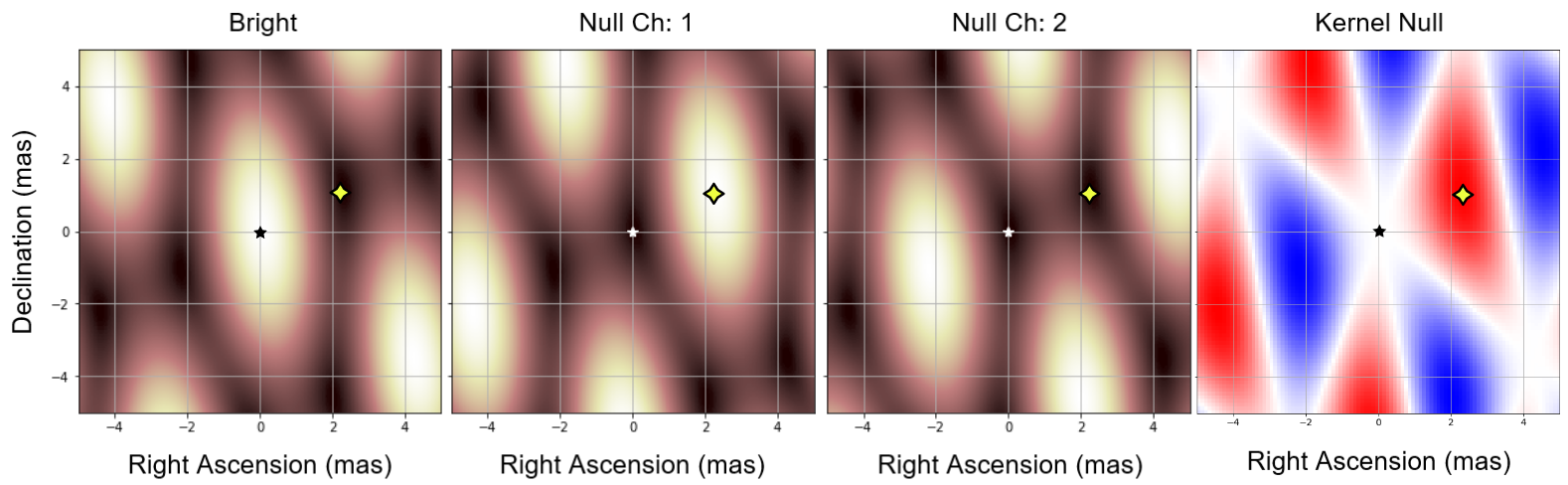}
\caption{The on-sky projection of the interferometer outputs assuming the use of three of the VLTI UTs at $\lambda = 1.55$~$\mu$m observing at zenith. Due to the interferometric beam combination inside the MMI the three output waveguides see different parts of the effective field of view. The on-axis star is seen by the bright output, but lies inside an interferometric null for the two nulled outputs. The synthetic binary companion (of contrast $10^2$) appears at a separation of 2.32~mas for this telescope arrangement, and is seen by all three channels by varying degrees. The response across the field of view of the kernel-null is shown on the far right. }
\label{fig:binaryprojection}
\end{figure*}

Our experimental setup and MC simulations allow us to study how our experimental kernel-nulls behave in the presence of increasing amount of residual piston. Figure~\ref{fig:KNvsRMS} plots the experimental results collected for induced piston residual RMS values up to 200~nm. We measure the expected increase in the distribution spread as the piston errors are increased, while remaining symmetrically distributed around the median. For data that is normally distributed the whiskers in Fig. \ref{fig:KNvsRMS} would contain $99.3\%$ of measured values, which for our experiments is true for RMS <25~nm. In this regime the distribution is dominated by sources of experimental error such as camera read-noise, leading to a statistically Gaussian distribution. Above this regime, the induced piston errors begin to dominate, with the measured distribution no longer being normally distributed (as seen in Fig \ref{fig:100nmkernel}). The increasing fat-tailed nature of the distribution for high piston errors can be seen in Figure \ref{fig:KNvsRMS} with an increasing number of measurements beyond the whiskers.   

In addition to random OPDs, the accuracy at which the co-phasing set-point has been determined (described in Section \ref{sec:opdzero}) will impact the experimental raw null values, imposing a new unknown floor value. This in turn will limit the self-calibrating capacity of the kernel-nuller, now affected by a consistent, albeit small, unknown bias.

The worst-case scenario would be to have an OPD set-point accuracy comparable to the RMS of the OPD random excursions. Working under the assumption of $100$~nm RMS OPD residuals, we will look at the evolution of the kernel-null bias over a $\pm 100$~nm offset range. For this, we rely on the MC simulation that produced the simulated distribution shown in Fig.~\ref{fig:100nmkernel}. We plot (see Fig.~\ref{fig:kappabias}) the evolution of the kernel-null median ($\kappa$) as a function of the OPD set-point error for input  $\#2$ and $\#3$, relative to input $\#1$. We can see that in the worst-case scenario, the kernel median bias can go up to $6 \times 10^{-4}$. The accuracy of our experimental co-phasing procedure is, in practice, better than $10$~nm and the resulting OPD set-point induced bias on our kernel-null median estimate ($\sim 10^{-5}$) is negligible in comparison to the effect of the bench drift outlined earlier. The level of sensitivity to this OPD co-phasing accuracy could be used as a criterion to decide between several types of kernel-nullers, recombining a different number of sub-apertures \citep{mikeLIFEKnull} and will be the object of future study as we move toward the characterization of a four-input kernel-nuller (Chingaipe et al. in prep).

Although the ultimate measurement uncertainty of $\kappa$ increases at larger piston excursions for the same number of frames, this series of experiments demonstrates that at least in this single on-axis source scenario, kernel-nulls retain the advertised self-calibrating property. To confirm that this is true in a more general scenario, we will have to use experiments including a simulated companion.


\subsection{Companion detection experiments}
\label{sec:results_companion}

To illustrate the use of kernel nulling in the presence of a companion, we created a binary-simulator acquisition sequence using our DM. This sequence interlaces and co-adds frames between two piston offset positions: the zero piston (as defined in Sec. \ref{sec:opdzero}) for the on-axis star, and a set of static piston values for the  off-axis companion. Frames alternating between these two set-points are acquired for every value of random piston errors, like was done in Sec. \ref{sec:results_kernel}. The off-axis frame is scaled by the desired contrast ratio, and added to the on-axis frame, simulating the observation of two mutually incoherent fluxes. This method of co-adding and scaling frames is not a perfect analogue of a spatially incoherent binary source as it also scales the associated noise in the frames of the companion signal. However, this only becomes important in the low-SNR regime where photon and background noise dominate.    
\begin{figure}[htp!]
\centering
\includegraphics[width=0.45\textwidth]{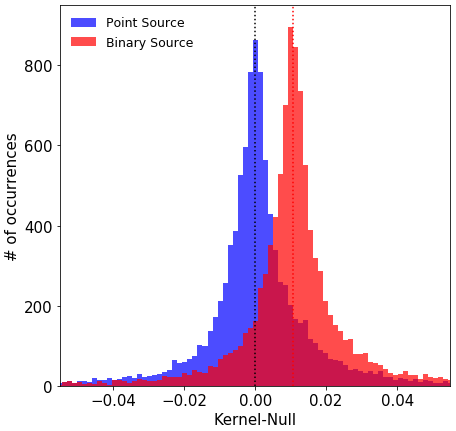}
\caption{ Measured kernel-null distributions under 100~nm RMS piston error for both an unresolved on-axis source and a synthetic binary source. The binary source contains the on-axis signal plus an additional incoherent companion signal with a contrast of $c=1\times10^{-2}$ and a separation of $2.32$~mas (see Fig. \ref{fig:binaryprojection}). The flux of the off-axis companion is interfered differently in the two null channels, causing an offset of the kernel-null distribution when compared to the point source. The measured mean of the binary source $\kappa_b = 1.06\times 10^{-2} \pm 8.1\times 10^{-4}$ is shown as the dotted red line.     }
\label{fig:kernelbinary}%
\end{figure}

We chose static piston offsets of 510 and -520~nm, respectively applied to inputs 2 and 3. This set of OPDs (marked by the blue diamond drawn on maps of Fig. \ref{fig:kscan}) corresponds to a point where the magnitude of the kernel signal is expected to be maximum. How these values map to the on-sky off-axis location of a companion depends on the particular configuration of sub-apertures that form the interferometric baselines. Figure~\ref{fig:binaryprojection} shows a possible on-sky mapping of these settings, on the VLTI long-baseline interferometer, using UT-2, 3, \& 4 observing at zenith, which would correspond to the presence of an off-axis companion at separation $\sim$2.3~mas. For a target in one of the nearby young associations at 140~pc, this separation would be equivalent to an orbital distance of $\sim 0.3$~AU.

This procedure was used to simulate the observation of a $1\times10^{-2}$ contrast companion, in the presence of piston residuals of 100~nm RMS. The respective kernel-null distribution recorded in this scenario is shown in Fig.~\ref{fig:kernelbinary}, alongside the distribution of a non-binary signal (unresolved on-axis source). The kernel-null mean for the binary signal was measured to be $\kappa_b = 1.06\times 10^{-2} \pm 8.1\times 10^{-4}$. The experimental uncertainty was determined by repeat measurements and is dominated by drift in our experimental setup. As the placement of the companion was chosen to be close to be where the kernel response is maximised, the measured value of the kernel-null mean directly infers the companion contrast. To infer the contrast of a companion at any other off-axis location would have to take into account the local response of the kernel-nuller. These experiments validate the main property of the nuller architecture devised by \cite{Martinache2018} and demonstrate that a kernel-nuller can advantageously be used to disentangle astrophysical signals from seeing induced leakage in a very direct manner.

\section{Discussion}

In this body of work we demonstrated experimentally that the previously theoretical concept of creating self-calibrated interferometric kernel-nulls is indeed possible. To our knowledge, this is the first physical demonstration of this kind of self-calibrated nulling, and while the detailed data-analysis tools for extracting the astrophysical kernel-null from noisy data are still being developed, these first results do confirm the advantages brought by the kernel-nuller architecture. The photonic implementation used for this work not only shows that advanced multi-aperture kernel-nulling interferometers can successfully be manufactured in integrated optics using publicly available commercial foundries, but also that it can be done so using more advanced photonic components such as MMIs. 

\begin{figure}[htp!]
\centering
\includegraphics[width=0.48\textwidth]{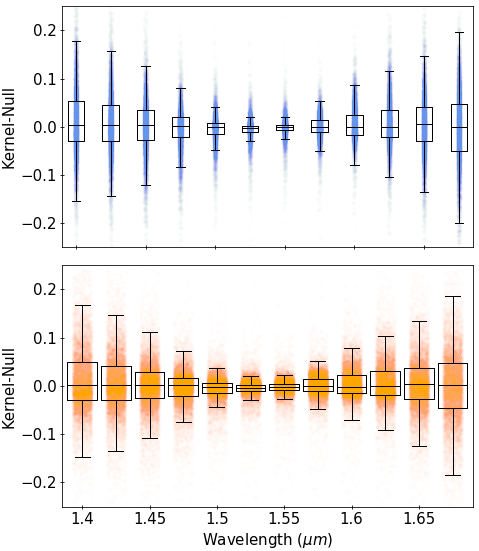}
\caption{ The measured kernel-null distribution under 100~nm RMS induced piston error across the astronomical H-band, for $12$~nm (top) \& $50$~nm (bottom) wide wavelength bins. The self-calibration performs the best at the MMI design wavelength ($1525-1575$~nm), then degrades as chromatic effects internal to the MMI de-tune the phase-matching. }
\label{fig:spectralKernel}%
\end{figure}

While the key results presented in Section \ref{sec:results} were for a spectrally narrow wavelength-band ($\Delta \lambda = 12$~nm) centered on $1.55$~$\mu$m for which the device was primarily designed, nothing fundamentally prevents its use over much of the astronomical H-band ($1.5-1.8$~$\mu$m), where the device remains single-mode and transmissive. Because the three nuller outputs are spectrally dispersed, our setup can be used to characterize the spectral behavior of this $3\times3$ MMI kernel-nuller. Two approaches are used here: either by broadening the spectral bandwidth that is binned into a single measurement to increase SNR, or by measuring a separate kernel-null for each wavelength (or some combination of both). The wavelength axis can provide benefits with an extra modelling constraint when fitting the contrast and angular separation parameters to a binary detection, as well as recording spectral information of any faint companion. In a practical sense the broadband performance of the kernel nuller is predominantly determined by the achromaticity of the MMI component. Our first MMI prototype was not fully optimized for broad-band use, so the self-calibration degrades as the wavelength deviates away from the central design wavelength of $1.55$~$\mu$m. Nevertheless, the kernel-null can still be extracted over much of the H-band as shown in Figure~\ref{fig:spectralKernel}. Here data was taken using the same method outlined in Section~\ref{sec:results_kernel}, but at different parts of the output spectrum, for two wavelength bin sizes ($12$~nm~\&~$50$~nm).

The scope of this work was limited to experimentally demonstrating kernel-nulling, thus an in-depth characterization of ultimate detection-limits and global device throughput over a broad wavelength band is left for future work. Any experimental validation of detection-limits is highly dependant on the  statistical uncertainty when measuring the median of the kernel-null distribution ($\sigma_{\kappa}$). Ensuring that the uncertainty is only due to the device response and not any external environmental effects requires the bench to be stable enough to obtain a $\sigma_{\kappa}$ of $\sim 10^{-5}$ (predicted by MC modelling). This is slightly beyond the capability of our bench in its current configuration, however can be achieved with active monitoring and control of injection flux to overcome the drift in the super-continuum source, and bonding of the output fibers.    

While the SiN platform used in this work was the best option available for low-cost prototyping of complex circuits, it might not be the case when creating final devices for on-sky use. The platform used is typically driven by the wavelength region of interest. In fact, the next generation of VLTI nulling instruments such as VIKiNG/Hi5 \citep{vikingSPIE2018} is proposed to work in the mid-IR around 4~$\mu$m wavelength, where the MMIs shown in this work are already being fabricated and tested on a Chalcogenide platform \citep{harry1, harry2}. Furthermore, as mentioned earlier, the same effect can be achieved using waveguide tri-couplers which have been manufactured by ULI, a platform already well characterized in an on-sky context. Nevertheless, this architecture in combination with the test-bench we have created is well placed to provide unparalleled experimental insights into this new interferometric technique.

Photonic kernel-nulling devices are not limited to the 3-beam architecture that was the focus of this work. Our group has started the characterization work of several 4-beam devices, closer to the design originally outlined by \cite{Martinache2018}, including a $4\times4$ design, consisting (like was the case here) of a single MMI, that is very similar in function, to the better known double-Bracewell design; as well as more complex designs featuring more outputs, made out of more complex photonic circuitry. Whereas much performance comparison and optimization work remains to be done, we can confidently suggest today that photonic solutions do provide an attractive alternative to the larger bulk-optics equivalents that have been previously explored \citep{Martin2012}, and will lead to the design of powerful miniature nulling beam-combiners well suited to ground-based as well as space-borne applications.

In an astrophysical context, the behaviour demonstrated here makes kernel-nulling an ideal candidate for use in the next generation of instrumentation for exoplanet science. Whether utilized on a long-baseline interferometer, or as part of a segmented single-aperture, the self-calibration shown here could enable nulling to overcome its sensitivity to residual piston errors, and aid in studying the inner parts of exoplanetary systems.

\section{Conclusions}

In this paper we experimentally demonstrate for the first time a successful creation of self-calibrated kernel-null observables for nulling interferometry. We achieved this through the use of a purpose-built photonic integrated device, containing a multimode interference coupler that creates one bright, and two nulled outputs when injected with three in-phase telescope beams. The interferometric transfer matrix created by the MMI produced the nulled outputs in a way that can be combined linearly to create a self-calibrated kernel-null. We use a broadband characterization bench with a deformable mirror to inject signals from both an unresolved point-source and a binary source analogue. We demonstrate the behaviour of the kernel-null for up to $200$~nm RMS wavefront error, and show the response for a $10^{-2}$ contrast binary companion at an angular separation of 2.32~mas under piston residuals of $100$~nm RMS. 

With the central self-calibrating feature of the kernel-nuller architecture now experimentally validated, we can confidently move on to work on more ambitious objectives: architectures that recombine more inputs and result in deeper nulls, relying on components and design choices that provide a stable null over broad bandpasses, that will enable the powerful spectroscopic nulling applications required for the characterization of the atmosphere of extrasolar planets with long baseline interferometry in space or on the ground. At the focus of a single telescope, the ability to produce reliable high-contrast observations, even in the presence of imperfect AO correction, makes kernel-nulling interferometry an interesting alternative to more conventional high-contrast imaging approaches. Even though it requires sparse apertures, one can still make good use of the full telescope pupil by combining several kernel-nullers operating in parallel on the light collected from several parts of a single telescope, resulting in a design with a near $100\%$ efficiency, an aggressive inner working angle, smaller than $\lambda/D$, possibly fully made out of photonics.


\begin{acknowledgements}
      The authors acknowledge the funding from the European Research Council (ERC) under the European Union's Horizon 2020 research and innovation program (grant agreement CoG - 683029). R. Laugier acknowledges the funding from the European Research Council (ERC) under the European Union's Horizon 2020 research and innovation program (grant agreement CoG - 866070).
\end{acknowledgements}

%
%

\bibliographystyle{aa} 
\bibliography{biblo}

\begin{thebibliography}{35}
\expandafter\ifx\csname natexlab\endcsname\relax\def\natexlab#1{#1}\fi

\bibitem[{Angel \& Woolf(1997)}]{Angel1997}
Angel, J. R.~P. \& Woolf, N.~J. 1997, The Astrophysical Journal, 475, 373

\bibitem[{{Angerhausen} \& {Quanz}(2021)}]{LIFE2021}
{Angerhausen}, D. \& {Quanz}, S. 2021, in European Planetary Science Congress,
  EPSC2021--284

\bibitem[{Baldwin {et~al.}(1986)Baldwin, Haniff, Mackay, \& Warner}]{BaldwinCP}
Baldwin, J.~E., Haniff, C.~A., Mackay, C.~D., \& Warner, P.~J. 1986, Nature,
  320, 595

\bibitem[{Beichman {et~al.}(2006)Beichman, Lawson, Lay, Ahmed, Unwin, \&
  Johnston}]{TPFIspie2006}
Beichman, C., Lawson, P., Lay, O., {et~al.} 2006, in Advances in Stellar
  Interferometry, ed. J.~D. Monnier, M.~Schöller, \& W.~C. Danchi, Vol. 6268,
  International Society for Optics and Photonics (SPIE), 245 -- 253

\bibitem[{{Bracewell}(1978)}]{6_bracewell}
{Bracewell}, R.~N. 1978, Nature, 274, 780

\bibitem[{Chingaipe {et~al.}(2022)Chingaipe, Cvetojevic, Matrinache, Laugier,
  Ławniczuk, Broeke, Ligi, N'Diaye, \& Mary}]{PeteSPIE2022}
Chingaipe, P., Cvetojevic, N., Matrinache, F., {et~al.} 2022, SPIE, in prep

\bibitem[{Cooney \& Peters(2016)}]{MMI_review}
Cooney, K. \& Peters, F.~H. 2016, Opt. Express, 24, 22481

\bibitem[{Cvetojevic {et~al.}(2021)Cvetojevic, Norris, Gross, Jovanovic,
  Arriola, Lacour, Kotani, Lawrence, Withford, \& Tuthill}]{dfly2021}
Cvetojevic, N., Norris, B. R.~M., Gross, S., {et~al.} 2021, Appl. Opt., 60, D33

\bibitem[{{Defr{\`e}re} {et~al.}(2016){Defr{\`e}re}, {Hinz}, {Mennesson},
  {Hoffmann}, {Millan-Gabet}, {Skemer}, {Bailey}, {Danchi}, {Downey}, {Durney},
  {Grenz}, {Hill}, {McMahon}, {Montoya}, {Spalding}, {Vaz}, {Absil}, {Arbo},
  {Bailey}, {Brusa}, {Bryden}, {Esposito}, {Gaspar}, {Haniff}, {Kennedy},
  {Leisenring}, {Marion}, {Nowak}, {Pinna}, {Powell}, {Puglisi}, {Rieke},
  {Roberge}, {Serabyn}, {Sosa}, {Stapeldfeldt}, {Su}, {Weinberger}, \&
  {Wyatt}}]{defrere2016}
{Defr{\`e}re}, D., {Hinz}, P.~M., {Mennesson}, B., {et~al.} 2016, The
  Astrophysical Journal, 824, 66

\bibitem[{Defrère {et~al.}(2018)Defrère, Ireland, Absil, Berger, Danchi,
  Ertel, Gallenne, Hénault, Hinz, Huby, Kraus, Labadie, Bouquin, Martin,
  Matter, Mennesson, Mérand, Minardi, Monnier, Norris, de~Xivry, Pedretti,
  Pott, Reggiani, Serabyn, Surdej, Tristram, \& Woillez}]{vikingSPIE2018}
Defrère, D., Ireland, M., Absil, O., {et~al.} 2018, SPIE, 10701, 223

\bibitem[{Goldsmith {et~al.}(2017{\natexlab{a}})Goldsmith, Cvetojevic, Ireland,
  \& Madden}]{harry1}
Goldsmith, H.-D.~K., Cvetojevic, N., Ireland, M., \& Madden, S.
  2017{\natexlab{a}}, Opt. Express, 25, 3038

\bibitem[{Goldsmith {et~al.}(2017{\natexlab{b}})Goldsmith, Ireland, Ma,
  Cvetojevic, \& Madden}]{harry2}
Goldsmith, H.-D.~K., Ireland, M., Ma, P., Cvetojevic, N., \& Madden, S.
  2017{\natexlab{b}}, Opt. Express, 25, 16813

\bibitem[{Guyon {et~al.}(2013)Guyon, Mennesson, Serabyn, \& Martin}]{Guyon2013}
Guyon, O., Mennesson, B., Serabyn, E., \& Martin, S. 2013, Publications of the
  Astronomical Society of the Pacific, 125, 951

\bibitem[{{Hanot} {et~al.}(2011){Hanot}, {Mennesson}, {Martin}, {Liewer},
  {Loya}, {Mawet}, {Riaud}, {Absil}, \& {Serabyn}}]{hanot2011}
{Hanot}, C., {Mennesson}, B., {Martin}, S., {et~al.} 2011, The Astrophysical
  Journal, 729, 110

\bibitem[{{Hansen} {et~al.}(2021){Hansen}, {Ireland}, {Ross-Adams}, {Gross},
  {Lagadec}, {Travouillon}, \& {Mathew}}]{tricoupler_aus}
{Hansen}, J.~T., {Ireland}, M.~J., {Ross-Adams}, A., {et~al.} 2021, arXiv
  e-prints, arXiv:2112.05017

\bibitem[{{Hansen} {et~al.}(2022){Hansen}, {Ireland}, \& {the LIFE
  Collaboration}}]{mikeLIFEKnull}
{Hansen}, J.~T., {Ireland}, M.~J., \& {the LIFE Collaboration}. 2022, arXiv
  e-prints, arXiv:2201.04891

\bibitem[{{Hinz} {et~al.}(1998){Hinz}, {Angel}, {Hoffmann}, {McCarthy},
  {McGuire}, {Cheselka}, {Hora}, \& {Woolf}}]{Hinz1998}
{Hinz}, P.~M., {Angel}, J. R.~P., {Hoffmann}, W.~F., {et~al.} 1998, Nature,
  395, 251

\bibitem[{Jennison(1958)}]{Jennison1958}
Jennison, R.~C. 1958, Monthly Notices of the Royal Astronomical Society, 118,
  276

\bibitem[{{Kaltenegger} {et~al.}(2003){Kaltenegger}, {Karlsson}, {Fridlund}, \&
  {Absil}}]{darwin}
{Kaltenegger}, L., {Karlsson}, A., {Fridlund}, M., \& {Absil}, O. 2003, in ESA
  Special Publication, Vol. 539, Earths: DARWIN/TPF and the Search for
  Extrasolar Terrestrial Planets, ed. M.~{Fridlund}, T.~{Henning}, \&
  H.~{Lacoste}, 459--464

\bibitem[{{Lacour} {et~al.}(2019){Lacour}, {Dembet}, {Abuter}, {F{\'e}dou},
  {Perrin}, {Choquet}, {Pfuhl}, {Eisenhauer}, {Woillez}, {Cassaing},
  {Wieprecht}, {Ott}, {Wiezorrek}, {Tristram}, {Wolff}, {Ram{\'\i}rez},
  {Haubois}, {Perraut}, {Straubmeier}, {Brandner}, \& {Amorim}}]{gravityFT}
{Lacour}, S., {Dembet}, R., {Abuter}, R., {et~al.} 2019, \aap, 624, A99

\bibitem[{Lacour {et~al.}(2014)Lacour, Tuthill, Monnier, Kotani, Gauchet, \&
  Labeye}]{Lacour2014}
Lacour, S., Tuthill, P., Monnier, J.~D., {et~al.} 2014, Monthly Notices of the
  Royal Astronomical Society, 439, 4018

\bibitem[{{Laugier} {et~al.}(2020){Laugier}, {Cvetojevic}, \&
  {Martinache}}]{romain2020}
{Laugier}, R., {Cvetojevic}, N., \& {Martinache}, F. 2020, \aap, 642, A202

\bibitem[{Martin {et~al.}(2012)Martin, Booth, Liewer, Raouf, Loya, \&
  Tang}]{Martin2012}
Martin, S., Booth, A., Liewer, K., {et~al.} 2012, Appl. Opt., 51, 3907

\bibitem[{Martinache(2010)}]{Martinache2010}
Martinache, F. 2010, The Astrophysical Journal, 724, 464

\bibitem[{Martinache \& Ireland(2018)}]{Martinache2018}
Martinache, F. \& Ireland, M.~J. 2018, Astronomy and Astrophysics, 619, 1

\bibitem[{{Martinod} {et~al.}(2021){Martinod}, {Norris}, {Tuthill}, {Lagadec},
  {Jovanovic}, {Cvetojevic}, {Gross}, {Arriola}, {Gretzinger}, {Withford},
  {Guyon}, {Lozi}, {Vievard}, {Deo}, {Lawrence}, \& {Leon-Saval}}]{GLINT2021}
{Martinod}, M.-A., {Norris}, B., {Tuthill}, P., {et~al.} 2021, Nature
  Communications, 12, 2465

\bibitem[{Martinod {et~al.}(2021)Martinod, Tuthill, Gross, Norris, Sweeney, \&
  Withford}]{tricoupler_MAM}
Martinod, M.-A., Tuthill, P., Gross, S., {et~al.} 2021, Appl. Opt., 60, D100

\bibitem[{{Mennesson} {et~al.}(2006){Mennesson}, {Haguenauer}, {Serabyn}, \&
  {Liewer}}]{6_PFN}
{Mennesson}, B., {Haguenauer}, P., {Serabyn}, E., \& {Liewer}, K. 2006, SPIE,
  6268, 626830

\bibitem[{{Norris} {et~al.}(2020){Norris}, {Cvetojevic}, {Lagadec},
  {Jovanovic}, {Gross}, {Arriola}, {Gretzinger}, {Martinod}, {Guyon}, {Lozi},
  {Withford}, {Lawrence}, \& {Tuthill}}]{norris2020}
{Norris}, B. R.~M., {Cvetojevic}, N., {Lagadec}, T., {et~al.} 2020, Monthly
  Notices of the Royal Astronomical Society, 491, 4180

\bibitem[{Roeloffzen {et~al.}(2018)Roeloffzen, Hoekman, Klein, Wevers, Timens,
  Marchenko, Geskus, Dekker, Alippi, Grootjans, van Rees, Oldenbeuving, Epping,
  Heideman, Wörhoff, Leinse, Geuzebroek, Schreuder, van Dijk, Visscher,
  Taddei, Fan, Taballione, Liu, Marpaung, Zhuang, Benelajla, \&
  Boller}]{lionix}
Roeloffzen, C. G.~H., Hoekman, M., Klein, E.~J., {et~al.} 2018, IEEE Journal of
  Selected Topics in Quantum Electronics, 24, 1

\bibitem[{Serabyn {et~al.}(2012)Serabyn, Mennesson, Colavita, Koresko, \&
  Kuchner}]{KIN2}
Serabyn, E., Mennesson, B., Colavita, M.~M., Koresko, C., \& Kuchner, M.~J.
  2012, The Astrophysical Journal, 748, 55

\bibitem[{{Serabyn} {et~al.}(2019){Serabyn}, {Mennesson}, {Martin}, {Liewer},
  \& {K{\"u}hn}}]{serabyn2019}
{Serabyn}, E., {Mennesson}, B., {Martin}, S., {Liewer}, K., \& {K{\"u}hn}, J.
  2019, Monthly Notices of the Royal Astronomical Society, 489, 1291

\bibitem[{Soldano \& Pennings(1995)}]{MMI_first}
Soldano, L. \& Pennings, E. 1995, Journal of Lightwave Technology, 13, 615

\bibitem[{Spalding {et~al.}(2022)Spalding, Defrère, \&
  Ertel}]{NullingReview2022}
Spalding, E., Defrère, D., \& Ertel, S. 2022, Physics Today, 75, 46

\bibitem[{Velusamy {et~al.}(2003)Velusamy, Angel, Eatchel, Tenerelli, \&
  Woolf}]{Velusamy2003}
Velusamy, T., Angel, R.~P., Eatchel, A., Tenerelli, D., \& Woolf, N.~J. 2003,
  European Space Agency, (Special Publication) ESA SP, 2003, 631

\end{thebibliography}

\end{document}